\newcounter{secnum}

\newcommand{\mysection}[1]{%
\vspace{1.25\baselineskip}
\stepcounter{section}
\stepcounter{secnum}
\centerline{\large\bf\thesecnum. #1}
\vspace{1pt}
}

\newcounter{subsecnum}[secnum]

\newbox\slashbox
\newdimen\slashwd
\newcommand{\slashed}[1]{%
\setbox\slashbox=\hbox{$#1$}%
\slashwd=\wd\slashbox%
\hbox to\slashwd{\hss/\hss}%
\llap{$#1$}}

\newcommand{\Deltaslash}{\slashed{\Delta}}

\newcommand{\pslash}{\slashed{p}}
\newcommand{\Pslash}{\slashed{P}}

\documentstyle[12pt]{article}

\begin{document}
 
\begin{titlepage}
\begin{center}
{\bf HOW TO RENORMALIZE A QUANTUM GAUGE FIELD THEORY}\\
{\bf WITH CHIRAL FERMIONS}\\
\end{center}

\begin{center}
Hung Cheng\\
Department of Mathematics, Massachusetts Institute of Technology\\
Cambridge, MA  02139, U.S.A.\\

\bigskip
and
\bigskip

S.P. Li\\
Institute of Physics, Academia Sinica\\
Nankang, Taipei, Taiwan, Republic of China\\
\end{center}

\vskip 3 cm

\begin{center}
{\bf Abstract}
\end{center}
 
We propose using the method of subtraction to renormalize quantum gauge theories with 
chiral fermions and with spontaneous symmetry breaking. 
The Ward-Takahashi identities derived from 
the BRST invariance in these theories are complex and rich in content.
We demonstrate how to use these identities to determine relationships among
renormalization constants of the theory and obtain the subtraction constants
needed for the renormalization procedure.  We 
have found it particularly convenient to adopt the Landau gauge throughout the scheme.  
The method of renormalization by subtraction enables
one to calculate physical quantities in the theory in the form of a renormalized
perturbation series which is unique and definite.  There is no ambiguity 
in handling the $\gamma_5$ matrix associated with chiral fermions.


\end{titlepage}

\newpage

\mysection{Introduction}

We present a scheme to renormalize quantum gauge field theories with
chiral fermions and with spontaneous symmetry breaking.  This scheme is
based on the method of subtraction with the aid of the Ward-Takahashi
identities.  The use of the Landau gauge is particularly helpful.

As we all know, a distinctive feature of
quantum field theories with spontaneous symmetry breaking is that the number of renormalized
parameters invariably exceeds that of the bare parameters. As a general
rule, a quantum field theory with an excessive number of renormalized
parameters is likely to be not renormalizable. Take, for example, the
theory of scalar QED.  It is well-known that, if we follow spinor QED to
the letter and introduce only two bare parameters, the bare charge and
the bare mass of the scalar meson, the corresponding quantum field
theory is not renormalizable. This is because there are three parameters
which must be renormalized, the third one being the $|\phi^4|$ coupling
constant. In order to make scalar electrodynamics renormalizable, one
additional parameter, the unrenormalized $|\phi^4|$ coupling
constant, must be introduced. It is notable that the Abelian-Higgs field
theory is an exception to the general rule. In this theory, there are
four bare constants in the boson sector: the bare coupling of the gauge
meson, the bare Yukawa coupling constant of the fermion, and the two bare
parameters in the Higgs potential. (There is also a parameter $\theta$ for the
mixing of the left-handed and right-handed components of the fermion.  This
parameter is finite and requires no renormalization.)  These four
bare constants generate more than four renormalized parameters: the
physical masses of the fermion, the gauge meson and the Higgs meson, the renormalized
coupling constants of the gauge meson, the renormalized Yukawa coupling
constants, and various renormalized 3-point and
4-point coupling constants of the scalar meson. While the number of
renormalized parameters exceeds that of the bare parameters, the
Abelian-Higgs theory is renormalizable. An important reason for this is
that the Ward-Takahashi identities put a constraint among some of these
parameters.

The contents of the Ward-Takahashi identities in a gauge field theory with
spontaneous symmetry breaking are complicated and require some care to
disentangle. In order to sidestep these complications, the method of
dimensional regularization has been invented[1,2]. With the use of
this method, the renormalized perturbation series automatically
satisfies the Ward-Takahashi identities, and one may be led to believe that there
is no more need to pay a great deal of attention to the identities which
have already been incorporated.

We consider such a belief misplaced. While the method of dimensional
regularization is convenient to use in explicit calculations, there are
three shortcomings associated with it. First, the definition of the matrix
$\gamma_{5}$ for a non-integral dimension is subjective and controversial up to now[3].
Therefore, the renormalization theory on the basis of dimensional
regularization alone remains incomplete.  Second, without explicitly exploring
the consequences of the Ward-Takahashi identities, the question of excessive
renormalized parameters remains unanswered.  Third, and perhaps most
important, the Ward-Takahashi identities contain rich and physical implications
which should be explored. Thus, ignoring the Ward-Takahashi identities under the
auspices of dimensional regularization is not an unmitigated blessing.
Indeed, it is not unlike throwing away the water with the baby in it.

We believe that the Ward-Takahashi identities are the centerpieces of
renormalization. Their contents should be extracted and utilized. The
method of dimensional regularization, on the other hand, should be
recognized as it is: a mathematical artifice which is helpful to use in
some cases---no more and no less.

We shall demonstrate our method of renormalization by the specific example 
of the Abelian-Higgs model.
The Langrangian density in the Abelian-Higgs model is given by
$$\begin{array}{ll}
L = & - \displaystyle \frac{1}{4} F_{\mu\nu} F^{\mu\nu} + (D_{\mu}\phi)^+ (D^\mu \phi) +
\bar{\psi}_{L}(i\partial\!\!\!/ - g_0 (1+\theta)V\!\!\!\!/) \psi_L

 + \bar{\psi}_R (i\partial\!\!\!/ - g_0 \theta V\!\!\!\!/) \psi_R \\ 
& - \sqrt{2} f_0 \phi
\bar{\psi}_L \psi_R - \sqrt{2}f_0 \phi^+ \bar{\psi}_R \psi_L + \mu^{2}_{0} \phi^{+}\phi - \lambda_0(\phi^{+}\phi)^2,
\end{array}
\eqno {(1)}
$$
with
$$
D_\mu \phi \equiv (\partial_\mu + ig_0 V_\mu)\phi, 
$$
$$
\psi_L \equiv \frac{1}{2}(1+ \gamma_5)\psi, 
$$
$$
\psi_R \equiv \frac{1}{2}(1- \gamma_5)\psi, 
$$
and
$$
F_{\mu\nu} \equiv \partial_\mu V_\nu - \partial_\nu V_\mu.
$$
In the above, $V_{\mu},\phi$, and $\psi$ are the gauge field, the complex scalar
field, and the fermion field, respectively, and $\theta, g_0, f_0, \mu_0$ and 
$\lambda_0$
are constants.  The subscript of the constants in
(1) signifies that these constants are bare.  As usual, we shall put
$$
\phi \equiv \frac{v_0 + H + i \phi_2}{\sqrt{2}}
$$
where
$
v_0 \equiv {\mu_0}/{\sqrt{\lambda_0}}.
$
We shall also call the bare mass of the gauge meson as
$
M_{0}\equiv g_0v_0,
$
and the bare mass of the fermion as 
$
m_0 \equiv f_0v_0.
$

We shall add to the Lagrangian (1) a gauge-fixing term and a ghost term.  Thus,
we consider the effective Lagrangian 
$$
L_{eff} \equiv L - \frac{1}{2\alpha}\ell^2 -
i(\partial_{\mu}\eta)(\partial^{\mu}\xi)+i \alpha M^2_0 \eta\xi + i \alpha g_0
M_0 \eta\xi H, 
\eqno{(2)}
$$
where
$$
\ell = \partial_{\mu} V^{\mu} - \alpha M_0 \phi_2,
$$
$\alpha$ is a constant, and $\xi$ and $\eta$ are ghost fields.

We shall first discuss the renormalization of the propagator of the chiral
fermion.  By Lorentz covariance, the 1PI amplitude of this propagator is 
of the form
$$
\delta m + ma(p^2) +m \gamma_5b(p^2)+c(p^2)\pslash+d(p^2)\pslash\gamma_5
\eqno{(3)}
$$
where 
$
\delta m \equiv m-m_0.
$
The unrenormalized propagator $S(p)$ is therefore given by
$$
S(p) = i\,\bigl[ \pslash(1-c)-m(1+a)-(bm+d\pslash)\gamma_5 \bigr]^{-1} . 
\eqno{(4)}
$$
The invariant amplitudes $b$ and $d$ are absent in QED, hence there are,
in the Abelian-Higgs theory with chiral fermions, two more invariant
functions which must be rendered finite by renormalization.

Fortunately the Abelian-Higgs theory remains renormalizable in spite of this,
as both of these amplitudes are finite.  To see this, we note that $S(p)$ by
assumption has a pole at
$
p_0 \equiv \sqrt{\vec{p}^2+m^2.}
$
The pole of $S(p)$ at $p_0=E$ comes from the integration of $<0| T \psi(x)
\bar{\psi}(y) |0>$ over very large
values of time, as the finite range of $(x_0-y_0)$ cannot
contribute a value of infinity.  By the adiabatic hypothesis, $\psi(x) \,
(\bar{\psi}(y))$ turns
into the out-field (in-field) as $x_0 \rightarrow \infty (y_0 \rightarrow
-\infty)$.  Since the in-field and the out-field are free bispinor fields of
mass $m$, it is straightforward to derive 
$$
c(m^2) = -a(m^2),
\eqno{(5a)}
$$
and
$$
b(m^2) = d(m^2) = 0,
\eqno{(5b)}
$$
which are the subtraction conditions needed for the divergent amplitudes
$a,b,c$ and $d$.

>From (4), we find that, when $p^2 \approx m^2$ and $\pslash$ is set to m,
$$
S(p) \approx \frac{2mi}{p^2-m^2}Z_{\psi}(m)
\eqno{(6)}
$$
where
$$
Z_\psi(m) \equiv \frac{1}{1-c(m^2)-2m^2[c\prime(m^2)+a\prime(m^2)]}.
\eqno{(7)}
$$
In deriving (6) and (7), we have utilized (5).  We shall define the renormalized
fermion propagator 
$
S^{(r)}(p) \equiv S(p)/Z_\psi(m) .
$

In a perturbative calculation, the 1PI amplitude for the fermion
propagator is linearly divergent.  As we know, a shift of the momentum
variable of a linearly divergent integral gives birth to a finite term.
Therefore, we may interpret a Feynman integral of the 1PI amplitude of
the fermion propagator as one of symmetric integration, with an unknown
additive constant arising from an undetermined amount of shift of the
momentum variables.  The symmetrically integrated amplitude is
logarithmically divergent.

There are two observations: (i) Because the amplitudes $a$ and $b$ in
(3) are multiplied by a factor of $m$, by power counting these
amplitudes are not linearly divergent.  Therefore, they do not contain
unknown additive constants, which appear only in the amplitudes $c$ and
$d$, (ii) There are no counter terms in the Lagrangian for the
amplitudes $d$ and $b$.  Since the amplitude $b$ does not contain an
unknown additive constant, it must be finite and must satisfy (5b) on
its own.  The amplitude $d$, on the other hand, is allowed an additive
constant contributed by linearly divergent integrals.  This constant is
determined by (5b).

The renormalized perturbation series for the renormalized fermion
self-energy 1PI amplitude can therefore be obtained as follows.  For a
graph which has no divergent subgraphs, we employ the Feynman rules to
obtain perturbatively the amplitudes $a,b,c,$ and $d$ with the coupling
constants and masses being the renomalized ones.  The divergent integrals are
symmetrically integrated after Feynman parameters are introduced.  The
amplitude $ma(p^2)+c(p^2)\pslash$ remains to be logarithmically
divergent, and is replaced by the subtracted amplitude
$$
 m \bigl[ a(p^2)-a(m^2) \bigr] + \bigl[ c(p^2)-c(m^2) \bigr] \pslash
 -2m^2 \bigl[ a'(m^2)+c'(m^2) \bigr] (\pslash-m).
\eqno{(8)}
$$
These subtractions are the same as the ones in QED.  They are so prescribed that
$S^{(r)}(p) \approx \displaystyle\frac{2mi}{p^2-m^2}$ near the mass shell with
$p\!\!\!/$ set to $m$.  The amplitude $b(p^2)$
requires no subtraction, while the amplitude $d(p^2)$ is replaced by the
subtracted amplitude
$
d(p^2)-d(m^2),
$
so that (5b) is satisfied.  For graphs with divergent subgraphs, we use the
BPHZ formalism[4].  We shall express the renomalized 1PI amplitude for the
fermion propagator as
$$
m a_r(p^2)+m\gamma_5b_r(p^2)+c_r(p^2)\pslash+d_r(p^2)\pslash\gamma_5.
\eqno{(9)}
$$

Next we turn to the Ward-Takahashi identity for three-point functions involving
fermions.  
We shall adopt the Landau gauge and take the limit $\alpha\rightarrow 0$.  In
this gauge, we have
$$
-
i\Delta_\mu\Gamma^{\mu}_{V\psi\bar{\psi}}(p^\prime,p)
-
m_0 Z\Gamma_{\phi_2 \psi\bar{\psi}}(p^\prime,p)
=
S^{-1}(p^\prime)\frac{1+2\theta+\gamma_5}{2}
-
\frac{1+2\theta-\gamma_5}{2}S^{-1}(p)
.
\eqno{(10)}$$
In (10), $p'$ and $p$ are the outgoing and incoming momenta
for the fermion, respectively, and 
$
\Delta=p^\prime-p.
$
Also,
$$
Z \equiv \displaystyle \frac{1}{v_0} < 0|(v_0 + H) |0> .
$$
The functions $\displaystyle\Gamma^{\mu}_{V\psi\bar{\psi}}$,
$\displaystyle\Gamma_{\phi_2\psi\bar{\psi}}$ and $S(p)$ are so defined that
their lowest-order terms in the unrenormalized perturbation
series are $\displaystyle\gamma^{\mu}\biggl[ (1+\theta)\frac{1+\gamma_5}{2}+ \theta
\frac{1-\gamma_5}{2} \biggr]$,$\displaystyle-i\gamma_5$ and
$\displaystyle\frac{i}{\pslash-m}$, respectively.
If we set $\Delta=0$ in (10), we get
$$
m_0
Z\Gamma_{\phi_2\psi\bar{\psi}}(p,p)=-\frac{1}{2}[\gamma_5S^{-1}(p)+S^{-1}(p)\gamma_5] .
\eqno{(11)}
$$
By Lorentz covariance, $\displaystyle\Gamma_{\phi_2\psi\bar{\psi}}$ is a
superposition of scalar amplitudes and pseudo-scalar amplitudes:
$$\begin{array}{ll} 
\Gamma_{\phi_{2}\psi\bar{\psi}}(p',p) &\equiv \displaystyle
\frac{1}{Z_{\phi_{2}\psi\bar{\psi}}(m^2, m^2, 0)} 
\biggl[ -i\gamma_5 F_0+F_1+i\gamma_5\Pslash
F_2+\Pslash F_3 \displaystyle +i\Deltaslash(1+\theta)\frac{1+\gamma_5}{2}
G_+\\ 
& + i\Deltaslash\theta
\displaystyle\frac{1-\gamma_5}{2}
G_- + (H_+ (1+\theta)
\displaystyle\frac{1+\gamma_5}{2}+H_{-}\theta\displaystyle\frac{1-\gamma_5}{2})(\Deltaslash\Pslash-\Pslash\Deltaslash) \biggr] ,
\end{array}
\eqno{(12)}
$$
where $F_i, G_{\pm},$ and $H_{\pm}$ are invariant amplitudes which
depend on $p^2,p'^2$ and $\Delta^2$, and $P\equiv \frac{1}{2}(p'+p)$.
Also, $F_0 (m^2, m^2, 0) = 1$ by definition.
By power counting, $F_1$ is logarithmically divergent, while $F_2,
F_3, G_{\pm}$ and $H_{\pm}$ are ultraviolet finite.  Substituting (12)
into (11), we get
$$
\frac{m_{0}Z}{Z_{\phi_2\psi\bar{\psi}}(m^2,m^2,0)}=m \bigl[ 1+a(m^2) \bigr] , 
\eqno{(13a)}
$$
and
$$
F_i(m^2,m^2,0)=0, \hspace{.50in} i = 1,2,3.
\eqno{(13b)}
$$

Equation (13b) with $i=1$ insures that the amplitude $F_1$ is actually
ultraviolet convergent.  Let us define the
renormalized coupling constant $f$ by
$$
f  \displaystyle \equiv f_0 \frac{\sqrt
{Z_{\phi_2}(0)}Z_\psi(m)}{Z_{\phi_2\psi\bar{\psi}}(m^2,m^2,0)} 
  =\frac{m}{v}[1+a(m^2)]Z_\psi(m^2),
\eqno{(14)}
$$
where the renormalized vacuum expectation value $v$ is given by[5]
$$
v\equiv v_{0}Z/\sqrt{Z_{\phi_{2}}(0)}.
$$
Eq. (14) can be written as
$$
f=\frac{m}{v}[1+a_r(m^2)].
\eqno{(15)}
$$
Thus the values of $m$ and $f$ are related.  Hence the renormalized Yukawa
coupling constant $f$ is finite as long as $m$ is finite, and vice versa.
Thus the Ward-Takahashi identities help make the theory renormalizable despite
the excessive number of renormalized parameters.

Let us next study the Ward-Takahashi identity (10) in the limit where 
$\Delta$ is infinitesimal but not zero.  We shall keep track of terms in this
identity which are linear in $\Delta$.  We remark that the point of subtraction
for $\Gamma_{V\psi\bar{\psi}}$ requires some care.  This is because
that in the Landau gauge adopted in our formalism,
$\Gamma_{V\psi\bar{\psi}}$ is infrared divergent if we set both $p^2$ and
${p'}^2$ to $m^2$, as $\phi_2$ is massless in the Landau gauge. 
Such a divergence is superficial, as all infrared divergent terms in
a physical scattering amplitude must cancel.  The point is that there is no
infrared divergence for the physical amplitudes evaluated in the 
$\alpha$-gauge with $\alpha \ne 0$.  Since the on-shell amplitudes are $\alpha$
independent[6], the on-shell amplitudes in the Landau gauge are 
infrared finite.  More precisely, at $p^2 = p'^2 = m^2$, the matrix elements $\bar{u}(p') 
\Gamma_{V\psi\bar{\psi}} u(p)$ are
infrared finite.  Nevertheless, some of the matrix elements of 
$\Gamma_{V\psi\bar{\psi}}$
do have infrared divergence where $p^2$ and $p'^2$ are both equal to $m^2$.
Thus the point of subtraction for $\Gamma_{V\psi\bar{\psi}}$ will be chosen 
to be at $p^2 = p'^2
= \Omega^2$, where $\Omega^2$ is not equal to $m^2$.  The physical 
amplitudes are $\Omega$-independent. 
By Lorentz covariance, $\Gamma^\mu_{{V}\psi\bar{\psi}}$ has
terms proportional to $\gamma^\mu, P^\mu$, and $\Delta^\mu$.  We shall
ignore terms proportional to the latter two in the Ward-Takahashi identity in 
the limit under consideration.  This is because in this identity 
 $\Gamma^\mu_{V\psi\bar{\psi}}$ is dotted with $\Delta_\mu$.
The terms $P^{\mu}$ and $\Delta^{\mu}$ in $\Gamma_{V\psi\bar{\psi}}^{\mu}$ 
can be ignored as $P^{\mu}$ 
dotted with $\Delta_{\mu}$ is equal to zero
while $\Delta_\mu\Delta^\mu$ is quadratic in $\Delta$.  Thus we shall replace
$\Gamma_{V\psi\bar{\psi}}$ in the Ward-Takahashi identity by
$$
\frac{1}{Z_{V\psi\bar{\psi}}(\Omega^2, \Omega^2, 0)} \biggl[
(\alpha_{+}+2\beta_{+} \Pslash)\gamma^\mu (1+\theta)
\frac{1+\gamma_5}{2}+(\alpha_{-}+2\beta_{-} \Pslash)\gamma^\mu\theta
\frac{1-\gamma_5}{2} \biggr] . 
\eqno{(16)}
$$
By power counting, the invariant amplitudes $\beta_+$ and $\beta_{-}$ are
ultraviolet finite.  We shall therefore pay attention to
$Z_{V\psi\bar{\psi}}$ and $\alpha_{-}$ only.  In the lowest-order,
$\alpha_{+}$ and $\alpha_{-}$ are both equal to $1$.  We shall define
$\alpha_{+}$ to be unity at the subtraction point.  Obviously, $\alpha_{-}$ is
not necessarily equal to unity at the point of subtract.  By substituting
$(16)$ into $(10)$ and equating the coefficients of
$-i\Deltaslash(1+\theta) \displaystyle\frac{1+\gamma_5}{2}$, we get
$$
\frac{Z_{\psi}(m)}
{Z_{V\psi\bar{\psi}}(\Omega^2,\Omega^2,0)}=1-c_r(\Omega^2)-d_r(\Omega^2)- m[1+a_r(m^2)]G_{+}(\Omega^2,\Omega^2,0).
\eqno{(17)}
$$

Eq. (17) shows that $Z_\psi/Z_{V\psi\bar{\psi}}$ is a finite number --- the
counterpart of $Z_2/Z_1=1$ in QED.  By equating coefficients of $-i\Deltaslash
\theta \displaystyle\frac{1-\gamma_5}{2}$ in (10), in the limit $\Delta$ infinitesimial and
$p^2={p'}^2= \Omega^2$, we get
$$
\frac{Z_{\psi}(m)}{Z_{V\psi\bar{\psi}}(\Omega^2,\Omega^2,0)}\alpha_{-}(\Omega^2,
\Omega^2,0)=1-c_r(\Omega^2)+d_r(\Omega^2)\\
-m[1+a_r(m^2)]G_{-}(\Omega^2, \Omega^2,0),
\eqno{(18)}
$$
which shows that $\alpha_{-}(\Omega^2,\Omega^2,0)$ is finite.  This also means
that the bare parameter $\theta$ should be chosen finite.

Using the rules of subtractions we have given in this section together
with the BPHZ formalism, one may perform renomalized perturbative
calculations to all orders.  The renormalized parameters in the Feynman
rules are required to obey relations such as (15).  The
presence of $\gamma_5$, the meaning of which is controversial in
dimensional regularization, presents no difficulty whatsoever in our
approach. 

One may also make use of the Ward-Takahashi identities for Green's functions
of bosons to show that the physical constants associated with bosons are 
finite[5].  Generalization to other gauge field theories such as that of
$SU(2) \times U(1) \times SU(3)$ is straightforward.   We shall demonstrate
in the following letter[7] an application  of this method by calculating the 
next-order triangular anomaly.

\newpage
\begin{center}
\large \bf {References}
\end{center}
 
\begin{enumerate}

\item {J. Ashmore, Lett. Nuovo Cimento \underline {4}, 289 (1972); C.G.
Bollini and J.J. Giambiagi, Phys. Lett \underline {40B}, 566 (1972).}

\item {G. 't Hooft and M. Veltman, Nucl. Phys. \underline{B44}, 189, (1972).}

\item {For a review of this issue, see, for example, Guy Bonneau,
International Journal of Modern Physics, Vol. 5, \underline{3831} (1990).}

\item {See, for example, Sec. 6.4 in S.J. Chang, Introduction to Quantum Field
Theory (World Scientific, 1990) for a simple discussion on the BPHZ
renormalization and also see the references therein.}

\item{H. Cheng and S.P. Li, Physical Masses and the Vacuum Expectation Value
of the Higgs field (to be published). }

\item{H. Cheng and E.C. Tsai, Chinese J. of Phys., Vol. \underline{25}, No. 1,
95 (1987), Phys. Rev. Lett., \underline{57}, 511 (1986); Phys. Rev. \underline
{D40}, 1246 (1989).}

\item{H. Cheng and S.P. Li, The radiative corrections of the Triangular Anomaly.
(submitted for publication)}

\end{enumerate}

\end{document}